\documentclass[11pt]{article}
\usepackage[T1]{fontenc}
\usepackage{a4wide}
\usepackage{graphics}

\makeatletter

\newcommand{\LyX}{L\kern-.1667em\lower.25em\hbox{Y}\kern-.125emX\spacefactor1000}

\usepackage{amssymb}
\makeatother

\begin{document}

July 1998\hfill{}DFUB-98-16 

\hfill{}

\bigskip{}
{\centering \textbf{\Large Scaling Functions in the Odd Charge Sector of Sine-Gordon/Massive
Thirring Theory}\Large \par}
\vspace{1cm}

{\centering G. Feverati\footnote{
E-mail: feverati@bo.infn.it
}, F. Ravanini\footnote{
E-mail: ravanini@bo.infn.it
} and G. Tak{\' a}cs\footnote{
E-mail: takacs@bo.infn.it
}\par}
\vspace{0.5cm}

{\centering \emph{INFN Sezione di Bologna - Dipartimento di Fisica}\\
\emph{Via Irnerio 46}\\
\emph{40126 Bologna, Italy}\par}

\begin{abstract}
A non-linear integral equation (NLIE) governing the finite size effects of excited
states of even topological charge in the sine-Gordon (sG) / massive Thirring
(mTh) field theory, deducible from a light-cone lattice formulation of the model,
has been known for some time. In this letter we conjecture an extension of this
NLIE to states with odd topological charge, thus completing the spectrum of
the theory. The scaling functions obtained as solutions to our conjectured NLIE
are compared successfully with Truncated Conformal Space data and the construction
is shown to be compatible with all other facts known about the local Hilbert
spaces of sG and mTh models. With the present results we have achieved a full
control over the finite size behaviour of energy levels of sG/mTh theory.\vspace{1cm}

\end{abstract}

\section{Introduction}

Finite size effects are widely recognized to play a major role in modern QFT
(Quantum Field Theory) \cite{luscher}. The concept of \emph{scaling functions,} encoding
the dependence of energy levels on the finite volume (Casimir effect), has shown
its practical convenience for a control of renormalization flows under change
of scale of many QFTs. Most of the progress made so far in computing the scaling
functions including non-perturbative effects have dealt with two-dimensional
QFT, in particular with integrable QFTs, where exact methods à la Bethe Ansatz
can be used to diagonalize Hamiltonians. A control of the behaviour of energy
levels of a two-dimensional integrable QFT under change of the finite volume
of a cylinder space-time geometry can lead not only to useful renormalization
flow information, but also serves as a basis for future developments in the
investigation of correlation functions \cite{magnoli-guida}.

A method of investigation that has been very fruitful, even for theories which
are not integrable, is the Truncated Conformal Space (TCS) approach \cite{yurov-zam}, which
is an intrinsically non-perturbative method, but has problems of principal nature,
coming from the fact that one does not have an analytic control of the spectrum,
and of practical nature, because to reach a certain precision in the resulting
energy levels one has sometimes to resort to very high truncation levels and
introduce enormous matrices to diagonalize. 

For integrable QFT, there also exist \emph{exact analytic} methods to compute
the finite size effects, like e.g. the Thermodynamic Bethe Ansatz (TBA) \cite{Al1}. Another
approach is the non-linear integral equation (NLIE) introduced as a continuum
limit of Bethe equations emerging from a light-cone lattice regularisation,
by Destri and De Vega \cite{ddv-92,ddv-95}. (Similar equations had already appeared in a different
condensed matter context in \cite{klumper}). These methods provide an impressive numerical
precision of the scaling functions, compared to the TCS one, and also give an
analytic control. The principal limitation of these exact methods was basically
that only the vacuum (and a few excited states degenerating with it at \( L\, \rightarrow \, \infty  \)) could
be obtained. However, progress of the last few years have opened the investigation
to excited states (see e.g. for the TBA approach refs. \cite{dorey-tateo,BLZ3}). In the 
Destri--De Vega approach,
the excited state equation was introduced in \cite{fioravanti-et-al} and later generalized to
include generic configurations of Bethe roots in \cite{ddv-97}. The correct quantisation
rule was established in \cite{noi1,noi2}, where it was shown that the resulting scaling functions
agree with numerical TCS data. 

The power of the NLIE method lies in the fact that it allows to describe all
excited states in a single framework for all values of the sG coupling constant,
and it can be used even in cases when the vacuum TBA is not known (in contrast
with the TBA approach of \cite{dorey-tateo,BLZ3}). It was argued in \cite{fioravanti-et-al} that a so-called \( \alpha  \)-twist following
the ideas of Zamolodchikov \cite{alyosha-polymer} gives an access to scaling functions in \( \Phi _{(1,3)} \) perturbations
of minimal models, the simplest examples of which include the scaling Yang-Lee
and the Ising model in zero magnetic field.

The NLIE still has a serious limitation, namely that only states lying in sectors
with even topological charge can be accessed. This restriction is an apparently
unavoidable feature of the light-cone lattice construction from which the NLIE
is deduced. However, it is of primary importance to master also the scaling
functions of the states with odd topological charge, as for example the single
soliton state in sG, or the single fermion state in mTh. In this letter we give
a conjecture of how the NLIE looks for such odd topological charge states, starting
from the results for the UV conformal dimensions computed in \cite{noi2}. The solutions
of the NLIE proposed for the odd sector are then compared to the TCS approach
for sG/mTh, i.e. perturbed \( c=1 \) CFT that we have developed in \cite{noi1,noi2}. Finally, the possibility
to twist the NLIE itself leads us to a full control of all periodic/antiperiodic
boundary conditions. This allows us to describe the differences between mTh
and sG that should be expected in the odd charge sectors, in full accordance
with the ideas of \cite{heretic}. 

Throughout the paper we adhere to the same notations as in \cite{noi1,noi2}. The reader is
invited to consult those papers (and references therein) as an introduction
to the concepts and general setup.

\section{UV analysis in the NLIE framework}

The NLIE describing the excited state spectrum of sG/mTh theory for even value
of the topological charge can be deduced from an inhomogeneous six-vertex model
put on a light-cone lattice and has the following form in the continuum limit
\begin{equation}
\label{nlie-cont}
\displaystyle Z(\vartheta )={\cal M}L\sinh \vartheta +g(\vartheta |\vartheta _{j})+2\Im m\displaystyle\int ^{\infty }_{-\infty }dxG(\vartheta -x-i\eta )\log \left( 1+(-1)^{\delta }e^{iZ(x+i\eta )}\right) \: ,
\end{equation}
where \( L \) is the spatial volume and \( {\cal M} \) is the soliton mass. We also introduce the
dimensionless volume parameter \( l={\cal M}L \). The function \( g(\vartheta |\vartheta _{j}) \) is the so-called \emph{source
term}, composed of the contributions from the holes, special objects and complex
roots (close and wide) which we call \emph{sources} and denote their positions
by the general symbol \( \{\vartheta _{j}\}=\{h_{k}\, ,\, y_{k}\, ,\, c_{k}\, ,\, w_{k}\} \) and is
\[
\displaystyle g(\vartheta |\vartheta _{j})=\sum ^{N_{H}}_{k=1}\chi (\vartheta -h_{k})-2\sum ^{N_{S}}_{k=1}\chi (\vartheta -y_{k})-\sum ^{M_{C}}_{k=1}\chi (\vartheta -c_{k})-\sum ^{M_{W}}_{k=1}\chi (\vartheta -w_{k})_{II}\: ,\]
where the function \( \chi  \) is the soliton-soliton phaseshift in sG theory, \( G \) is its
derivative up to a factor of \( 2\pi  \), and the index \( II \) denotes the second determination
necessary for wide roots (for details see \cite{noi2}). The positions of the sources are
related to the \emph{counting function} \( Z(\vartheta ) \) by the Bethe Ansatz quantisation conditions
\[
Z(\vartheta _{j})=2\pi I_{j}\, \, ,\, \, I_{j}\in \mathbb {Z}+\frac{1-\delta }{2}\, \, ,\]
where \( I_{j} \) are the Bethe quantum numbers. There is a so-called \emph{counting equation}

\begin{equation}
\label{counting}
N_{H}-2N_{S}=2S+M_{C}+2\theta (\pi -2\gamma )M_{W}\, \, ,
\end{equation}
relating the number of the different roots/holes to \( S \), which is the spin of
the XXZ chain appearing in the lattice regularisation and takes integer values
for the Bethe states. The sG topological charge \( Q \) turns out to be \( Q=2S \).

Once the solution for the counting function \( Z(\vartheta ) \) is known, the energy can be computed
from the formula
\begin{equation}
\label{energy}
\begin{array}{rl}
\displaystyle E & ={\cal M}\displaystyle\sum ^{N_{H}}_{j=1}\cosh h_{j}-2{\cal M}\displaystyle\sum ^{N_{S}}_{j=1}\cosh y_{j}\\
 & -{\cal M}\displaystyle\sum ^{M_{C}}_{j=1}\cosh c_{j}+{\cal M}\displaystyle\sum _{j=1}^{M_{W}}(\cosh w_{j})_{II}\\
 & -{\cal M}\displaystyle\int ^{\infty }_{-\infty }\displaystyle\frac{dx}{2\pi }2\Im m\left[ \sinh (x+i\eta )\log (1+(-1)^{\delta }e^{iZ(x+i\eta )})\right] \: ,
\end{array}
\end{equation}
with a similar expression for the momentum.

We start our investigation by recalling some facts about the UV analysis of
the spectrum of the NLIE, i.e. the consideration of the so called \emph{kink
equation} and its consequences. The UV limit of sG theory is \( c=1 \) conformal field
theory (CFT) with a compactification radius \( R \) which is related to the six-vertex
anisotropy \( \gamma  \) by
\begin{equation}
\label{compact_radius}
R^{-1}=\sqrt{2\left( 1-\displaystyle\frac{\gamma }{\pi }\right) }\, \, ,
\end{equation}
and it is often convenient to introduce the parameter\footnote{
\( p>1 \) (i.e. \( 0<\gamma <\pi /2 \)) corresponds to the repulsive while \( p<1 \) (i.e. \( \pi /2<\gamma <\pi  \)) to the attractive regime
of sG theory. The \( n \)th breather threshold is at \( p=1/n \).
}
\[
p=\frac{\pi }{\gamma }-1\, \, .\]
The primary fields are vertex operators \( V_{(n,m)} \) with conformal weights

\[
\Delta _{(n,m)}^{\pm }=\frac{1}{2}\left( \frac{n}{R}\pm \frac{1}{2}mR\right) ^{2}\]
where \( m\in \mathbb {Z} \) is the winding number (which is conserved in the off-critical theory
and is identified with the sG topological charge \( Q \)) and \( n \) is the field momentum.
The relation (\ref{compact_radius}) means that we identify the perturbing operator with \( V_{(1,0)}+V_{(-1,0)} \).

We recall some of the results of ref. \cite{noi2}. In the UV limit \( l\, \rightarrow \, 0 \) the sources can be
classified into three types: their position can remain finite (``central''), or
they can move towards the two infinities as \( \pm \log \frac{2}{l} \) (``left/right movers''). We denote
the number of right/left moving holes by \( N^{\pm }_{H} \) and similarly we introduce the numbers
\( N^{\pm }_{S} \), \( M^{\pm }_{C} \) and \( M^{\pm }_{W} \).

We introduce the \emph{partial spins} \( S^{+},\: S^{-} \) by the definition: 
\[
S^{\pm }=\frac{1}{2}[N_{H}^{\pm }-2N_{S}^{\pm }-M_{C}^{\pm }-2M^{\pm }_{W}\theta (\pi -2\gamma )]\, \, .\]
The expression for the conformal weights is
\begin{equation}
\label{delta}
\Delta _{\pm }=\pm \left( I^{\pm }_{H}-2I^{\pm }_{S}-I^{\pm }_{C}-I^{\pm }_{W}\right) +\displaystyle\frac{\Sigma _{\pm }}{2\pi }+\displaystyle\frac{\omega ^{2}_{\pm }}{16\pi ^{2}(1-\gamma /\pi )}\: ,
\end{equation}
where

\begin{equation}
\label{plateau_solution}
\omega _{\pm }=\pm 2(\pi -2\gamma )\left( S-2S^{\pm }\right) -2(\pi -\gamma )(\delta +2k_{\pm })\, \, ,
\end{equation}
\( k_{\pm } \) are integers determined by the condition \( -\pi \leq \omega _{\pm }\leq \pi  \),

\begin{equation}
\label{sigmapm}
\Sigma _{\pm }=-4S^{\pm }(S-S^{\pm })\displaystyle\frac{\pi -2\gamma }{1-\gamma /\pi }+2\pi q^{\pm }_{W}\, \, ,
\end{equation}
and we introduced the following notation for the sums of quantum numbers of
left/right moving sources of the different types
\[
I^{\pm }_{H}=\sum ^{N^{\pm }_{H}}_{j=1}I_{h_{j}}^{\pm }\, ,\, I^{\pm }_{C}=\sum ^{M^{\pm }_{C}}_{j=1}I_{c_{j}}^{\pm }\, ,\, I^{\pm }_{W}=\sum ^{M^{\pm }_{W}}_{j=1}I_{w_{j}}^{\pm }\, \mathrm{and}\, I^{\pm }_{S}=\sum ^{N^{\pm }_{S}}_{j=1}I_{y_{j}}^{\pm }\]
(\( q_{W}^{\pm } \) is an integer or half-integer depending on the configuration of the wide
roots, which is best calculated case by case). 

From (\ref{delta}) we obtain the following expressions:
\begin{equation}
\label{UV-limit}
\begin{array}{c}
m=2S\\
2n=(\delta +2S+M_{sc})\: \bmod 2
\end{array}
\end{equation}
where \( M_{sc} \) is the number of selfconjugate wide roots. Since \( S\in \mathbb {Z} \) we have access only
to sectors with even value of the topological charge \( Q=m \). In \cite{noi2}, the selection
rule for sG states was shown to be

\[
(\delta +M_{sc})\: \bmod 2=0\, \, .\]

Now let us consider putting \( S \) half-integer in (\ref{UV-limit}) and remember that in order
to have the sG with periodic boundary conditions we need \( n\in \mathbb {Z} \). Thus we obtain the
rule
\begin{equation}
\label{rule-sg}
(\delta +M_{sc})\: \bmod 2=1.
\end{equation}

From (\ref{counting}) we see that if \( S\in \mathbb {Z}+\displaystyle\frac{1}{2} \) then \( N_{H} \) must be odd (given that the number of close
complex roots \( M_{C} \) is always even). Therefore to describe such states we put an
odd number of hole sources in the NLIE. This extension of the equation \emph{cannot}
be derived from the light-cone lattice approach, since the latter only allows
for even number of holes. In the following we will take it as a conjecture and
give arguments in support. 

Let us now present examples of UV dimensions computed from the NLIE for some
simple states. For the state with one hole with quantum number \( I=0 \), using \( \delta =1 \) we
obtain the conformal dimensions
\[
\Delta ^{\pm }=\frac{1}{8R^{2}}\, \, ,\]
corresponding to the vertex operator \( V_{(0,1)} \). This operator can be identified as the
UV limit of the one-soliton field as it is well known \cite{heretic}. 

Consider now the case of three holes, with quantum numbers \( I_{1}=1 \), \( I_{2}=0 \) and \( I_{3}=-1 \). The conformal
dimension we obtain corresponds to the vertex operator \( V_{(0,3)} \) as expected. If we
choose other values of quantum numbers different from this ``minimal'' choice,
we obtain descendents of vertex operators \( V_{(n,3)} \), where \( n\in \mathbb {Z} \) depends on the number of
holes moving to the left and to the right.

As a last example, let us take three holes with quantum numbers \( I_{1}=1 \), \( I_{2}=0 \) and \( I_{3}=-1 \),
but we add a close pair with quantum number \( 0 \) for both complex roots. We obtain
the conformal weights
\[
\Delta ^{\pm }=\frac{1}{8R^{2}}+1\, \, ,\]
which is the first nonchiral descendent of \( V_{(0,1)} \).

Using the above results, we formulate our conjecture as follows: the sG states
with odd value of topological charge correspond to taking an odd number of holes
and choosing the quantisation rule \( \delta  \) in accordance with (\ref{rule-sg}). In the next section
we shall see that using this rule we get scaling functions which are in excellent
agreement with numerical results obtained from the \( c=1 \) Truncated Conformal Space
(TCS) method.

\section{Numerical comparison between the NLIE and TCS}

In this section we present the numerical results which give further support
to the conjecture on the identification of sG states with odd topological charge.
In particular, we compare the NLIE prediction to numerical results obtained
from the TCS method for the lowest energy levels.

\begin{figure}
{\centering \resizebox*{0.8\textwidth}{0.3\textheight}{\includegraphics{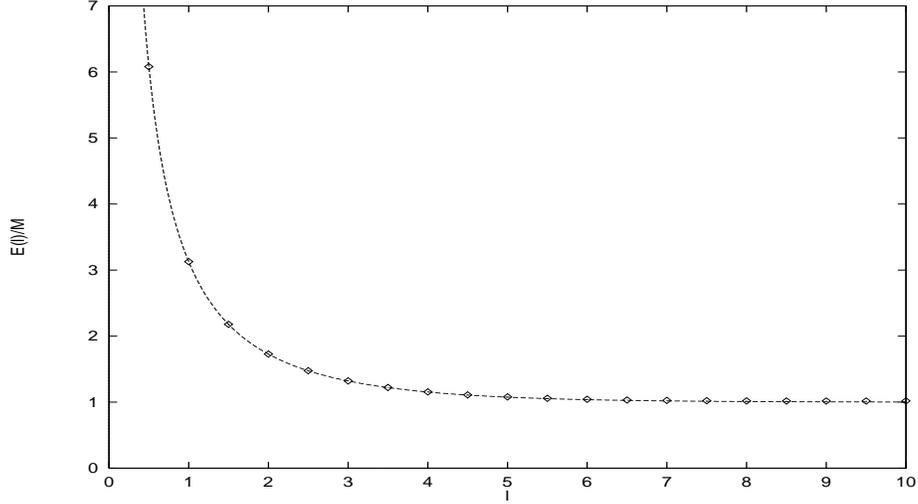}} \par}

\caption{\small \label{onesoliton_fig} Comparison of the results for the one-soliton state coming NLIE and TCS at
\protect\( p=\frac{2}{7}\protect \). The continuous line is the NLIE result, while the diamonds represent the
TCS data.}
\end{figure}

We start with the one-soliton state, which is conjectured to be a one-hole configuration
with integer quantisation. Figure \ref{onesoliton_fig} shows the comparison graphically, while
table \ref{onesoliton_table} gives an idea about the numerical magnitude of the difference. Note
that the deviations are extremely small for small values of \( l \) and they grow
with the volume, exactly as expected for truncation errors.  \begin{table}
{\centering \begin{tabular}{|c|c|c|c|}
\hline 
l&
NLIE&
TCS&
Relative deviation \\
\hline 
\hline 
.5&
6.080571&
6.08062&
0.000008\\
\hline 
1&
3.126706&
3.12685&
0.00005\\
\hline 
1.5&
2.177411&
2.17791&
0.0002\\
\hline 
2&
1.727224&
1.72776&
0.0003\\
\hline 
2.5&
1.475004&
1.47593&
0.0006\\
\hline 
3&
1.320353&
1.32168&
0.001\\
\hline 
4&
1.153188&
1.15548&
0.002\\
\hline 
5&
1.075376&
1.07908&
0.003\\
\hline 
\end{tabular}\par}

\caption{\small \label{onesoliton_table} Numerical comparison of the energy levels predicted by the NLIE to the TCS
data for the case depicted in figure \ref{onesoliton_fig}. The NLIE data are exact to the precision
shown.}
\end{table} 

\begin{figure}
{\centering \resizebox*{0.8\textwidth}{0.3\textheight}{\includegraphics{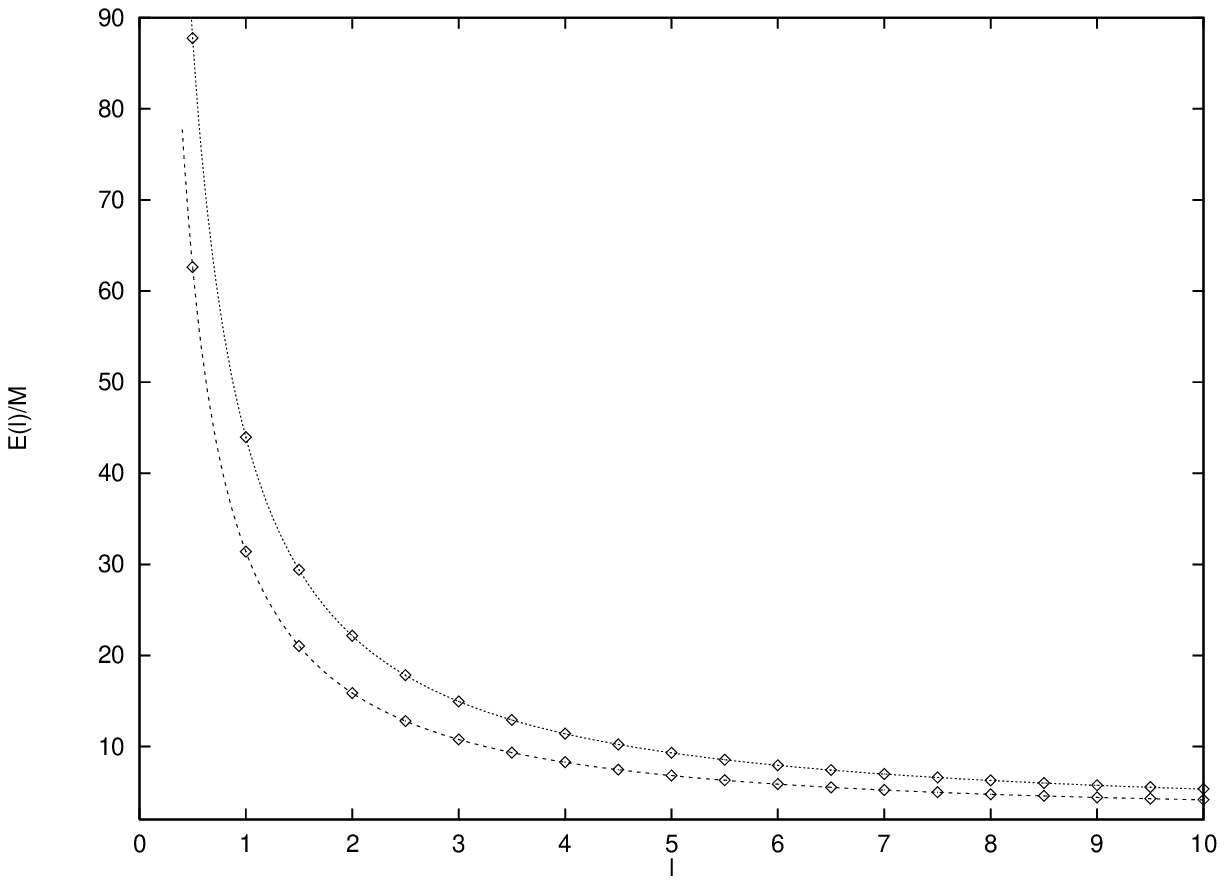}} \par}

\caption{\small \label{threesoliton_fig}The energies as a function of the volume for first two three-soliton states
coming from NLIE and TCS at \protect\( p=\frac{2}{7}\protect \). The continuous lines are the NLIE result, while
the diamonds represent the TCS data.}
\end{figure}

Next we take the two lowest energy levels corresponding to states of three solitons
with the same charge (i.e. three holes with quantum numbers \( 0,\pm 1 \) and \( 0,\pm 2 \), respectively).
Here we only present the energy plot (figure \ref{threesoliton_fig}). The numerical deviations are
exactly of the same magnitude as in the case of the one-soliton state.

The value of the coupling constant at which the above data are presented lies
in the attractive regime where the TCS method is free of UV divergences \cite{noi1,noi2} and
therefore we can compare absolute energy levels, after subtracting the linear
bulk energy term

\[
\frac{E_{bulk}}{M}=Bl,\, \, \, B=-\frac{1}{4}\tan \frac{\pi p}{2}\, \, ,\]
which is predicted by TBA \cite{alyosha-mass} and also by the NLIE \cite{ddv-95}, from the TCS data.

\begin{figure}
{\centering \resizebox*{0.8\textwidth}{0.3\textheight}{\includegraphics{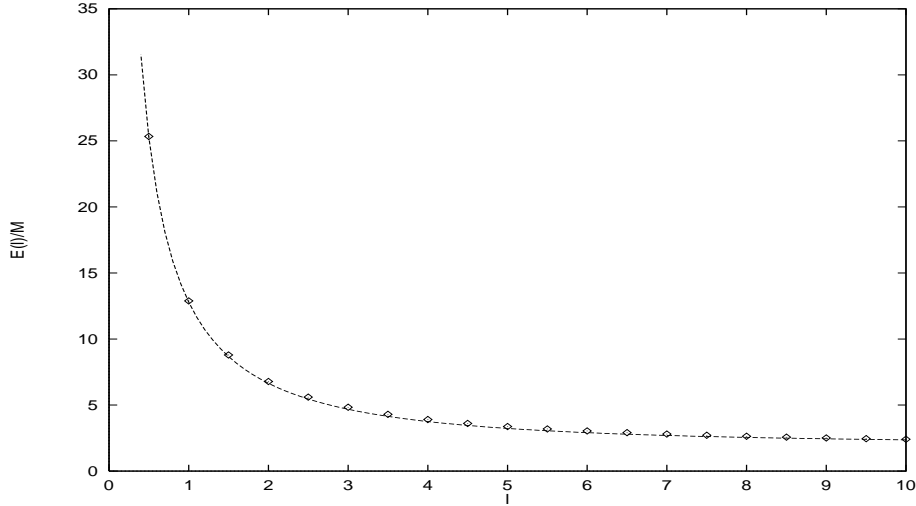}} \par}

\caption{\small \label{cpair3h_fig} The NLIE prediction for the state with three holes (with quantum numbers
\protect\( -1,\, \, 0\protect \) and \protect\( 0\protect \)) and a complex pair, at \protect\( p=1.5\protect \), compared to TCS data. Both are normalized
to the ground state of the \protect\( Q=1\protect \) sector, i.e. the one-soliton state, to get free
of the UV divergent part of the TCS data.}
\end{figure}

The state with three holes and one complex pair can be most easily investigated
in the repulsive regime, since it changes its nature in the attractive regime
where the complex roots behave in a more complicated way. The result is depicted
in figure \ref{cpair3h_fig}. As in the repulsive regime the TCS is UV divergent, one can only
compare relative energy levels, and as we are in the sector with unit topological
charge, the natural way is to normalize the energy levels to the one-soliton
state, which is the ground state of the \( Q=1 \) sector. Although this cannot be directly
seen from figure \ref{cpair3h_fig}, it is clear from the numbers presented in table \ref{cpair3h_table} that the
truncation errors are much larger even after subtracting the UV divergences.
The same phenomenon was observed in our earlier works \cite{noi1,noi2}.\begin{table}
{\centering \begin{tabular}{|c|c|c|c|}
\hline 
l&
NLIE&
TCS&
Relative deviation \\
\hline 
\hline 
.5&
25.27656&
25.3398&
0.0025\\
\hline 
1&
12.79207&
12.8854&
0.007\\
\hline 
1.5&
8.674948&
8.79114&
0.01\\
\hline 
2&
6.647771&
6.78091&
0.02\\
\hline 
2.5&
5.455906&
5.60033&
0.026\\
\hline 
3&
4.680766&
4.83188&
0.03\\
\hline 
4&
3.751729&
3.90497&
0.04\\
\hline 
5&
3.231486&
3.37598&
0.045\\
\hline 
\end{tabular}\par}

\caption{\small \label{cpair3h_table} Numerical comparison of the energy levels predicted by the NLIE to the TCS
data for the case depicted in figure \ref{cpair3h_fig}.}
\end{table}

\section{Scaling functions in the massive Thirring model}

Here we formulate another conjecture concerning the finite size energy levels
of the massive Thirring model. Of the two quantum numbers \( n,\, \, m \) specifying a vertex
operator \( V_{(n,m)} \), \( m \) can be identified with the sG topological charge and takes integer
values, while \( n \) is not conserved in the massive theory. Generally, one can introduce
twisted boundary conditions, for which 
\[
n\in \mathbb {Z}+\frac{\nu }{2\pi }\, \, ,\]
where the parameter \( \nu  \) lies between \( 0 \) and \( 2\pi  \). Such boundary conditions can be
realised by adding a constant term \( \alpha  \) to the right hand side of the NLIE which
is related to \( \nu  \) as
\[
\nu =\alpha -\pi \delta \, \bmod \, 2\pi \, \, .\]
Here we will need only the cases \( \nu =0 \) or \( \pi  \), which are realised by taking \( n\in \mathbb {Z}/2 \). For
convenience, we divide the Hilbert space into four sectors, given by

\[
\begin{array}{l}
{\cal H}_{I}=\bigoplus \{{\cal F}_{(n,m)}\, :\, \, n\in \mathbb {Z},\, \, m\in 2\mathbb {Z}\}\, \, ,\\
{\cal H}_{II}=\bigoplus \{{\cal F}_{(n,m)}\, :\, \, n\in \mathbb {Z},\, \, m\in 2\mathbb {Z}+1\}\, \, ,\\
{\cal H}_{III}=\bigoplus \{{\cal F}_{(n,m)}\, :\, \, n\in \mathbb {Z}+\displaystyle\frac{1}{2},\, \, m\in 2\mathbb {Z}+1\}\, \, ,\\
{\cal H}_{IV}=\bigoplus \{{\cal F}_{(n,m)}\, :\, \, n\in \mathbb {Z}+\displaystyle\frac{1}{2},\, \, m\in 2\mathbb {Z}\}\, \, ,
\end{array}\]
where \( {\cal F}_{(n,m)} \) denotes the Fock module corresponding to the vertex operator \( V_{(n,m)} \). According
to \cite{heretic}, the sG Hilbert space (for periodic boundary conditions) can be identified
with \( {\cal H}_{I}\oplus {\cal H}_{II} \) while that of mTh is \( {\cal H}_{I}\oplus {\cal H}_{III} \). Note that these two spaces are related to each
other by twisting the boundary conditions. The corresponding two algebras are two different
maximal closed local operator algebras, which are essentially unique (see comments
at the end of the section) at generic values of the compacitification radius
\( R \). \( {\cal H}_{IV} \) never appears as a local sector in sG/mTh theory. Using the results from
the UV limit of the NLIE (\ref{UV-limit}), we see that the massive Thirring model is selected
by (putting \( \alpha =0 \)) 
\begin{equation}
\label{mTh_rule}
\delta =M_{sc}\, \bmod \, 2\, .
\end{equation}
As an example, let us consider the one-hole state with half-integer quantisation
and quantum number \( I=\displaystyle\frac{1}{2} \). Using the general formulae we compute the UV dimensions
\[
\Delta ^{\pm }=\frac{1}{2}\left( \frac{1}{2R}\pm \frac{R}{2}\right) ^{2}\, \, ,\]
which corresponds to the vertex operator \( V_{(1/2,1)} \). If we choose instead \( I=-\frac{1}{2} \), we obtain
the UV limit \( V_{(-1/2,1)} \). These are exactly the positive charge components of the (Dirac)
fermion field in the \( c=1 \) CFT and so we can identify the one-hole configuration
with the one-fermion state of the massive Thirring model. One can also compute
the dimensions of more general configurations and verify that they describe
states in \( {\cal H}_{I}\oplus {\cal H}_{III} \).

The iterative solution of the NLIE presents serious difficulties when the arrangement
of the source terms is not symmetric under \( \vartheta \, \rightarrow \, -\vartheta  \). The number of steps generally
increases when \( l \) decreases. This can be prevented by making a re-antisymmetrization
of the counting function computed at each step, which however can be done only
for symmetric source configurations. For mTh states of odd charge, the source
configurations are never symmetric. In the case of the one-fermion state, the
iteration slows down too much to reach values of \( l \) less than \( 3.0 \). The solution
of this problem may necessitate the introduction of some fundamentally new numerical
algorithm to solve the NLIE and prevents us from presenting a numerical comparison
to TCS data of such mTh states for the time being. The results available for
regions of \( l \) where the iteration converges fast enough (see table \ref{1fermion_table}) show an
agreement with TCS similar to the one presented for sG states in the previous
section (cf. table \ref{onesoliton_table}).

\begin{table}
{\centering \begin{tabular}{|c|c|c|c|}
\hline 
l&
NLIE&
TCS&
Relative deviation\\
\hline 
\hline 
3.0&
1.549935&
1.55081&
0.0006\\
\hline 
3.5&
1.415153&
1.41657&
0.001\\
\hline 
4.0&
1.321588&
1.32357&
0.0015\\
\hline 
4.5&
1.254597&
1.25722&
0.002\\
\hline 
5.0&
1.205454&
1.20881&
0.003\\
\hline 
\end{tabular}\par}

\caption{\small \label{1fermion_table} The comparison between the NLIE results for the energy of the one-fermion
state and the TCS data for the range \protect\( 3.0<l<5.0\protect \). The coupling constant is \protect\( p=\frac{2}{7}\protect \) and the hole
quantum number was chosen \protect\( I=\frac{1}{2}\protect \).}
\end{table} 

Our selection rule (\ref{mTh_rule}) takes into account that, as described in our previous
paper, we redefined the source term for the self-conjugated roots (see section
3.2 of \cite{noi2}). If one keeps the form of the source terms as they are derived from
the lattice, one can summarize the selection rules for the even and the odd
sector of mTh model in the following simple form:
\[
\delta =0\, \bmod \, 2\, \, .\]
With the same convention, the selection rule for the sine-Gordon model becomes

\[
\delta =2S\, \bmod \, 2\, \, .\]

Let us make a remark on the uniqueness of the maximal local operator algebra.
The two operator algebras correspoding to the bosonic theory sG and the fermionic
theory mTh
\[
\begin{array}{l}
{\cal A}_{b}=\{V_{(n,m)}:\, \, n\in \mathbb {Z},\, m\in \mathbb {Z}\}\\
{\cal A}_{f}=\{V_{(n,m)}:\, \, n\in \mathbb {Z},\, m\in \mathbb {Z}+n/2\}
\end{array}\]
are the only ones at a generic value of the compactification radius \( R \) in the
sense that any other maximal local operator algebras can be mapped into them
by redefining \( R \) by multiplication with an integer \cite{heretic}. However, the identification 
of the perturbation
operator is not unique: it can be taken to be any of \( V_{k,0}+V_{-k,0} \) with \( k \) integer. This
would change the value of \( R \) in (\ref{compact_radius}) by a factor of \( k \). If we map these theories
back to perturbations by \( V_{1,0}+V_{-1,0} \) by redefining \( R \) to the value in (\ref{compact_radius}), we obtain the
following other local operator algebras:
\[
\begin{array}{l}
{\cal A}^{(k)}_{b}=\{V_{(n/k,km)}:\, \, m\in \mathbb {Z},\, n\in \mathbb {Z}\}\\
{\cal A}^{(k)}_{f}=\{V_{(n/k,km)}:\, \, m\in \mathbb {Z},\, n\in \mathbb {Z}+m/2\}
\end{array}\]
This is the complete list of possible local theories as \( k \) runs over all positive
integers. Note that sG/mTh corresponds to choosing \( k=1 \). Another interesting theory
is the one defined by \( {\cal A}^{(2)}_{b} \) with the corresponding Hilbert space 
given by \( {\cal H}_{I}\oplus {\cal H}_{IV} \). At the
Kosterlitz-Thouless point \( \gamma =0 \) where the perturbation becomes marginal this describes
the current-current perturbation of a level-\( 1 \) \( SU(2) \) WZNW model, which is nothing
else than the SU(2) Gross-Neveu theory. Note that the states with odd topological
charge are absent from the local sector. This is due to the fact that the elementary
excitation is a ``kink'' with fractional Lorentz spin \( 1/4 \). To describe the kink
state, created by the operator \( V_{\pm 1/4,\pm 1} \) in the UV, one has to twist the odd sector
NLIE by \( \alpha =\pm \pi /2 \). This sort of nonlocality is a general property of the models based
on \( {\cal A}^{(k)}_{b,f} \) which is why they were called ``kink'' theories in \cite{heretic}.

\section{Conclusions}

In this letter we have proposed a generalization of the NLIE to the odd charge
sectors of sG and mTh models. The conjecture has been tested against the data
coming from a TCS diagonalization of the \( c=1 \) perturbed CFT Hamiltonian that is
expected to give a description of sG/mTh models. The agreement is excellent
and gives strong evidence for our conjecture, which is also supported by the
fact that the proposal is the most obvious and natural generalization of the
even sector NLIE. Further evidence is given by the UV analysis reproducing correctly
the conformal dimensions of the states to be found in such sectors, and in particular
by the successful identification of the single sG soliton state as well as the
two components of the mTh fermion.

It is interesting to notice that the NLIE framework together with this proposal
allows for a complete treatment of both sG and mTh Hilbert spaces. The two models,
according to the interpretation of ref.\cite{heretic}, correspond to two subspaces of the
same space, identified in the UV limit by the action of two different maximal
local subalgebra of conformal operators. One leads to a counting of states described
by the modular invariant partition function, while the other corresponds to
the \( \Gamma _{2} \)-invariant one. mTh model corresponds to sG with twisted boundary conditions
and vice-versa. All these features can now be controlled also away from UV criticality,
thanks to the NLIE approach that we have in a sense ``completed'' with the present
extension to odd charge sectors.

Of course one could think to investigate different local UV operator algebras,
identifying other models, e.g. the algebras \( {\cal A}_{b,f}^{(k)} \). These are controllable on the
level of the NLIE by choosing appropriately the twist parameter \( \alpha  \) implementing
boundary conditions. Thus one could explore, e.g., the formulation of the deformed
\( SU(2) \) Gross-Neveu model which is the QFT having \( {\cal A}^{(2)} _{b} \) 
as local operator algebra away from the Kosterlitz-Thouless point. 
Other interesting QFTs arise at special (``rational'') values
of the radii, like the \( \Phi _{(1,3)} \) perturbations of minimal models. Notice that in the
language of ref. \cite{alyosha-mass}, all these models are just ``projections'' onto different local
sectors, of the same Hilbert space of sG model on the cylinder, if \( \alpha  \)-sectors
are meant to be included in the Hilbert space from the beginning.

It would be interesting, as a further development, to propose a modified light-cone
lattice formulation of sG/mTh, leading to Bethe equations that allow the treatment
of odd number of holes. A proposal in this respect has been put forward by Destri
and Segalini \cite{destri-segalini}, in order to avoid the nonlocality present in 
the original light-cone
lattice approach \cite{ddv-87}. This nonlocality is manifested both in the lattice Hamiltonian
and in the construction of the Thirring fermion field on the lattice. The fermionic
light-cone lattice model has Bethe equations that allow for odd number of holes
and look very similar to the ones derived from the ``traditional'' light-cone
approach. However, since in this framework the fermion field is represented
locally on the lattice, there appears the well-known doubling of fermionic species.
The doublers are shown to decouple from the physical fields in the continuum
limit of the Hamiltonian. In order to give a derivation of the NLIE starting
from the fermionic formalism it is necessary to follow this decoupling in the
Bethe equations themselves, which is out of the scope of this paper. We would
just like to point out that the rule of half-integer Bethe quantisation in the
mTh model is in complete accord with their fermionic lattice model.\\

\textbf{Acknowledgements -} We are indebted to C. Destri for useful discussions.
This work was supported in part by NATO Grant CRG 950751, by European Union
TMR Network FMRX-CT96-0012 and by INFN \emph{Iniziativa Specifica} TO12. G.
T. has been partially supported by the FKFP 0125/1997 and OTKA T016251 Hungarian
funds.


\begin{thebibliography}{10}
\bibitem[1]{luscher}M. Lüscher, \emph{Comm. Math. Phys.} \textbf{104} (1986) 177-206.\\
M. Lüscher, \emph{Comm. Math. Phys.} \textbf{105} (1986) 153-188.
\bibitem[2]{magnoli-guida}R. Guida and N. Magnoli, \emph{Phys. Lett.} \textbf{B411} (1997) 127-133, hep-th/9706017.
\bibitem[3]{yurov-zam}V.P. Yurov and A.B. Zamolodchikov, \textit{Int.J.Mod.Phys.} \textbf{A5} (1990)
3221-3246. 
\bibitem[4]{Al1}Al.B. Zamolodchikov, \textit{Nucl. Phys.} \textbf{B342} (1990) 695-720.
\bibitem[5]{ddv-92}C. Destri and H.J. De Vega, \textit{Phys. Rev. Lett.} \textbf{69} (1992) 2313-2317. 
\bibitem[6]{ddv-95}C. Destri and H.J. De Vega, \textit{Nucl. Phys.} \textbf{B438} (1995) 413-454,
hep-th/9407117.
\bibitem[7]{klumper}A. Klümper and P.A. Pearce, \textit{J. Stat. Phys.} \textbf{64} (1991) 13; \\
A. Klümper, M. Batchelor and P.A. Pearce, \textit{J. Phys.} \textbf{A24} (1991)
3111.
\bibitem[8]{dorey-tateo}P. Dorey and R. Tateo, \emph{Nucl. Phys.} \textbf{B482} (1996) 639-659, hep-th/9607167.
\\
P. Dorey and R. Tateo, \emph{Nucl. Phys.} \textbf{B515} (1998) 575-623, hep-th/9706140. 
\bibitem[9]{BLZ3}V.V. Bazhanov, S.L. Lukyanov and A.B. Zamolodchikov, \emph{Nucl. Phys.} \textbf{B489}
(1997) 487-531, hep-th/9607099. 
\bibitem[10]{fioravanti-et-al}D. Fioravanti, A. Mariottini, E. Quattrini and F. Ravanini, \textit{Phys. Lett.}
\textbf{B390} (1997) 243-251, hep-th/9608091.
\bibitem[11]{ddv-97}C. Destri and H. De Vega, \textit{Nucl. Phys.} \textbf{B504} (1997) 621-664,
hep-th/9701107.
\bibitem[12]{noi1}G. Feverati, F. Ravanini and G. Takács: \emph{Truncated Conformal Space at c=1,
Nonlinear Integral Equation and Quantisation Rules} \emph{for Multi-Soliton
States,} preprint DFUB-98-04, hep-th/9803104, \emph{Physics Letters} \textbf{B}
in press. 
\bibitem[13]{noi2}G. Feverati, F. Ravanini and G. Takács: \emph{Nonlinear Integral Equation and
Finite Volume Spectrum of Sine-Gordon Theory,} preprint DFUB-98-10, hep-th/9805117.
\bibitem[14]{alyosha-polymer}Al. B. Zamolodchikov, \emph{Nucl. Phys.} \textbf{B432} (1994) 427-456, hep-th/9409108.
\bibitem[15]{heretic}T. Klassen and E. Melzer, \emph{Int. J. Mod. Phys}\textbf{. A8} (1993) 4131-4174,
hep-th/9206114.
\bibitem[16]{alyosha-mass}Al.B. Zamolodchikov, \textit{Int. J. Mod. Phys.} \textbf{A10} (1995) 1125-1150. 
\bibitem[17]{destri-segalini}C. Destri and T. Segalini, \emph{Nucl. Phys.} \textbf{B455} (1995) 759, hep-th/9506120.
\bibitem[18]{ddv-87}C. Destri and H.J. De Vega, \emph{Nucl. Phys.} \textbf{B290} (1987) 363-391. 
\end{thebibliography}
\end{document}